# Results of the first user program on the Homogenous Thermal Neutron Source HOTNES (ENEA / INFN)

A. Sperduti[1,*], M. Angelone[2], R. Bedogni[3], G. Claps[2], E. Diociaiuti[3], C. Domingo[4], R. Donghia[3], S. Giovannella[3], J.M. Gomez-Ros[5], L. Irazola-Rosales[6], S. Loreti[2], V. Monti[7], S. Miscetti[3], F. Murtas[3], G. Pagano[2], M. Pillon[2], R. Pilotti[2], A. Pola[8], M. Romero-Expósito[4], F. Sánchez-Doblado[5], O. Sans-Planell[9], A. Scherillo[10], E. Soldani[3], M. Treccani[4], and A. Pietropaolo[3]

[1] Department of Physics and Astronomy, Uppsala University, SE-751 05 Uppsala, Sweden
[2] ENEA, Department of Fusion and Technologies for Nuclear Safety and Security, Frascati, Italy
[3] Istituto Nazionale di Fisica Nucleare, Frascati National Laboratories, Frascati, Italy
[4] Universitat Autonoma de Barcelona UAB, Physics Department, 08193 Bellaterra, Spain
[5] CIEMAT, Av. Complutense 40, 28040 Madrid, Spain
[6] Departamento de Fisiología Médica y Biofísica, Universidad de Sevilla, Spain
[7] INFN Sezione di Torino, via Pietro Giuria 1, 10125 Torino, Italy
[8] Politecnico di Milano, Energy department, via La Masa 34, 20156 Milano, Italy
[9] Universitat Rovira i Virgili Carrer de l`Escorxador, 43003 Tarragona, Spain
[10] Science and Technology Facilities Council ISIS facility, Chilton, Didcot, Oxfordshire, 0Q11, UK

## Abstract

The HOmogeneous Thermal NEutron Source (HOTNES) is a new type of thermal neutron irradiation assembly developed by the ENEA-INFN collaboration.

The facility is fully characterized in terms of neutron field and dosimetric quantities, by either computational and experimental methods. This paper reports the results of the first "HOTNES users program", carried out in 2016, and covering a variety of thermal neutron active detectors such as scintillators, solid-state, single crystal diamond and gaseous detectors.

**Keywords**: Thermal neutron source, neutron detectors

*Corresponding author: andrea.sperduti@physics.uu.se

## 1. Introduction

The HOmogeneous Thermal NEutron Source (HOTNES) [1, 2] (figure 1) was developed at the ENEA Frascati Research Centre [3], in the framework of INFN-ENEA collaboration, following a new moderator design conceived by Bedogni et al. [4]. HOTNES relies on a $^{241}$Am-B neutron source with strength $3.5\times10^6$ s$^{-1}$ (nominal value). A cylindrical 5 mm thick lead shield attenuates the 59.5 keV photons from $^{241}$Am. The source is located at the bottom of a large cylindrical cavity delimited by polyethylene walls, floor and cover. This cavity has diameter 30 cm and height 70 cm. A polyethylene shadow bar (10 cm in diameter and 20 cm in height) prevents fast neutron to directly reach the samples to be irradiated.

The useful irradiation volume is delimited on bottom by the shadow bar, and on top by a removable 5 cm thick polyethylene cover. The useful volume has diameter of 30 cm and 40 cm height (figure 2). The effects of the shadow bar and of the cavity walls combine in such a way that the thermal neutron fluence is nearly uniform across the so called iso-fluence disks which are parallel to the cavity bottom. Iso-fluence disks are identified by the height (Z) from the cavity bottom. The thermal neutron fluence rate at HOTNES reference irradiation plane (Z=50 cm) is 758±16 cm$^{-2}$s$^{-1}$ (in terms of *sub-cadmium cut off fluence rate in the Westcott convention* [5]), representing about 85% of the overall neutron fluence rate. The neutron field characterization was already presented in the dedicated paper here referred as Ref. [1].

The thermal field is roughly isotropic when the 5 cm thick polyethylene cover is in place, while if it is removed, a nearly parallel beam is achieved.

HOTNES was designed as a user facility for multi-purpose thermal neutron testing. Thus, a number of customers, owning different types of thermal neutron sensitive devices, were convened in ENEA for a first testing campaign. The following devices were tested:

- Li(Eu) glass scintillation detector (Li-glass) [6];
- Cadmium Zinc Telluride (CZT) solid state detector [7];
- Thermal Neutron Rate Detector (TNRD) [8];
- Single Crystal Diamond detector (SCD) [9];
- Gas Electron Multiplier (GEM) gaseous detector [10];
- Pure CsI scintillation detectors.

## 2. Experimental devices

This section introduces the detectors used in the campaign and their results.

Some small-sized detectors (Li-glass and CZT) were used to map HOTNES irradiation volume in terms of thermal neutron field. This volume (see figure 2), defined in height from the top of the shadow bar till the polyethylene cover, was scanned by varying the measurement position over both height (from Z=30 cm to Z=60 cm) and radius (from R=0 to R=14 cm).

The volume mapping was done in terms of the quantity $q(Z, R) = \dfrac{N(Z, R)}{N(50, 0)}$, i.e. the ratio of the count rate measured by the detector under test at a given position (Z, R) inside the HOTNES cavity to that at the reference position (Z=50, R=0). The quantity $q(Z, 0) = \dfrac{N(Z, 0)}{N(50, 0)}$ represents the ratio of the detector count rate measured at the central position of a given irradiation plane normalized to that measured at the central position of the reference plane.

For other detectors, punctual measurements at HOTNES reference irradiation position were performed for calibration purposes.

### 2.1. Li-glass scintillation detector

The Li-glass was purchased from Saint Gobain Glass [11]. The size of scintillator is 15 mm × 15 mm area and thickness 0.62 mm. Thermal neutron are detected by the $^6$Li(n,α)$^3$He reaction (Q=4.78 MeV) occurring inside the glass. Due to the considerable height of the associated photomultiplier tube (PMT) that was about 25 cm, the HOTNES polyethylene cover was removed to map irradiation planes Z=50 cm and 60 cm (figure 1). Lower planes (Z=30 cm and 40 cm) were mapped with and without the polyethylene cover.

The PMT was biased at +1100 V by means of a Caen HV N1470 and connected to a preamplifier, ORTEC 113 [12]. The signal then is processed by the ORTEC 474 Timing Filter Amplifier [13]. The output signal was processed by a PC board-based Multichannel Analyser (MCA), ORTEC 927 [14]. Referring to figure 3, the "neutron signal" was derived as the integral of the spectrum above channel 120.

Figures 4 and 5 show, respectively, the trend of $q(Z,R)$ and $q(Z,0)$ for the Li-glass. As far as figure 4 is concerned, it can be noted that at Z=50 cm the trend features a decrease of about 10% in going from R=0 cm to R=12 cm radial position. The same consideration can be done for the Z=60 cm plane. The lowest plane (Z=30 cm) shows the typical trend with a sudden count decrease in the region up to about R=6 cm due to the effect of the shadow bar, while for the Z=40 cm, a slightly different trend is found with an increasing intensity from R=0 cm to R=8 cm radial position of about 8% - 9%. Although the general trend may be considered consistent with respect to the TNPD (Thermal Neutron Pulsed Detectors) [15] measurements, nevertheless the big dimensions of the Li-glass detector likely induce perturbation of the neutron field at the LiF crystal measurement position. TNPD are manufactured at INFN-LNF by evaporation-based deposition technique of $^6$LiF powder 30 μm thick and a maximum depletion layer of 0.3 mm (at full depletion). The TNPD has an active area of 1 cm$^2$, 4 mm thick and is based on a silicon p-i-n diode. In figure 5, $q(Z,0)$ measured with Li-glass is compared with that measured with calibrated TNPD in the same conditions. The agreement is quite satisfactory within experimental uncertainties.

### 2.2. Cadmium Zinc Telluride solid state detector

CZT detectors, mostly used in X-ray spectroscopy, are ternary semiconductor compounds (Cd$_{1-x}$Zn$_x$Te), the x values ranging between 5% and 13% with a corresponding energy gap ranging between 1.53 and 1.48 eV. In these measurements the thermal neutron radiative capture on $^{113}$Cd (12.22% abundance) was exploited to detect thermal neutrons [6]. The CZT was purchased from EURORAD and the crystal had size 5×5×5 mm$^3$ [16]. The device was biased by means of a power HV SO16 EURORAD. The output signal was sent to a spectroscopy amplifier ORTEC 570 [17] and then processed by a PC board-based Multichannel Analyser (MCA). Each measurement lasted about 900 s, setting an electronic threshold for pulse height spectra recording at 60 keV. Total gain and shaping time of the spectroscopic amplifier were set at 30 and 2 μs, respectively. Neutron counts for each measurement were determined by subtracting the fitted gamma background beneath the 558.46 keV peak and then integrating around the cadmium peak. The gamma background was fitted using the function $I(E)=\alpha E^\beta$, with $\alpha$ and $\beta$ fitting parameters.

Figure 6 shows as an example the pulse height spectrum recorded by the CZT as placed at Z=50 cm and R=0 cm in the HOTNES irradiation cavity.

The results obtained with CZT detector for $q(Z,R)$ and $q(Z,0)$ are shown in figures 7 and 8, respectively. The detector count rate results to sensibly vary along the radial position (figure 7). The plane at Z=50 cm shows a change in the count rate of about 15% from R=0 cm to R=14 cm. The count rate at Z=40 cm is quite different as compare to the reference plane, with a difference in the count rate which is about 5% at R=0 cm and reaches 20% for 5 cm<R<10 cm. The Z=60 cm plane shows a high homogeneity (≈ 89%) and a count rate lower than 25% respect to the reference plane.

The discrepancies between the trend of $q(Z, R)$ and $q(Z, 0)$ measured with CZT and those obtained with Li-glass (and TNPD) are likely due to the gamma background subtraction used to extract neutron count rate as explained before when discussing figure 6.

### 2.3. Thermal Neutron Rate Detector (TNRD)

The TNRD detector was developed by Bedogni et al. [15] in the framework of the NESCOFI@BTF and NEURAPID projects (Scientific Commission V, INFN-LNF, Italy). The detector is based on a low-cost commercial solid-state device made sensitive to thermal neutrons through a customized $^6$LiF deposition layer. Its active area is 1 cm$^2$ and its overall dimensions are 1.5 cm × 1 cm × 0.4 cm. Its output is a DC voltage, which is proportional to the thermal neutron fluence rate (for this reason the device is called "rate detector"). This signal is amplified in a low-voltage electronics module especially developed by the project team. The output signal is sent to a programmable ADC (NI USB-6218 BNC) controlled by a PC through a LabView application.

TNRD linearly responds to thermal neutron fluence rates from $10^2$ up to $10^6$ cm$^{-2}$ s$^{-1}$. TNRDs have been largely used in measuring the dosimetric quantities associated to parasitic neutrons during radiotherapy with electron accelerators [18] .

In the current work, six TNRDs were individually calibrated in terms of sub-cadmium cut off fluence rate in the *Westcott convention*, using HOTNES reference irradiation plane (Z=50 cm). The calibration procedure consists in three stages (see figure 9):

1- The detector is located on the irradiation plane and is covered with borated rubber (attenuation factor about 300) for about 600 seconds. The corresponding output voltage constitutes the "left baseline", mainly due to the "dark" current in the electronics;

2- The borated rubber is removed, exposing the detector to the thermal beam for about 1000 seconds;

3- The detector is covered again with borated rubber for about 600 seconds, and the corresponding "right baseline" is recorded.

The net neutron signal, due to thermal neutrons (in the order of $10^{-2}$ V), is obtained by subtracting the average baseline (left and right) from the average voltage during the thermal neutrons exposure. This is divided by the thermal fluence rate to get the detector calibration factor. Table 1 reports the calibration factors measured in HOTNES for six different detectors. Uncertainties are in the order of 4% due to the combination of the following sources: uncertainty in the thermal fluence rate (2%), measurement statistics (< 3%), repeatability in detector positioning (2%). Even for this type of detector, HOTNES proved to be a useful testing tool, in terms of intensity as well as positioning and general set up.

### 2.4. GEM-based neutron detector

The triple-GEM detector is micro-pattern gas detector based on the Gas Electron Multiplier (GEM) [10]. A GEM is made of a thin kapton foil (50 µm), a copper cladding (5 µm) on each side, perforated with a high surface density of holes. Each hole has a bi-conical structure with external (internal) diameter of 70 µm (50 µm) and a pitch of 140 µm. The GEM is the basic building block of a new generation of gas detectors, in particular it is used as charge multiplication stage. The detector used in this work, consists of three GEM foils piled-up and sandwiched between two electrodes, a cathode and an anode. The triple-GEM configuration allows reaching higher detector gains before the appearance of discharges, without requiring too high voltage on each single GEM foil. The cathode is about 10×10 $cm^2$ area and the gaps of the drift region, of the inter-GEM region and of the induction region are 3, 1, 2, and 1 mm, respectively. The readout anode is divided in a matrix of 12×12 square pads, each one having a side of 8 mm. Read-out electronics is based on a set of 8 CARIOCA cards and FPGA mother board which reads the pads in digital mode with the possibility to set also a threshold level to cut electronic noise or very weak signals [19]. In order to use this detector for thermal neutrons, the cathode has been realized using an boron deposited aluminium plate. A 1.5 µm thick layer of $B_4C$ with $^{10}B$ enrichment was deposited by the Linköping University by means of sputtering technique [20]. The main acquisition parameters set during this experiment were: 1) the total voltage gain HV=960 V, provided by means of a HVGEM module [21], corresponds to a gain of about 100 and 2) the threshold level TL=1400 mV, which results to be just above the electronic noise. The GEM detector was used to perform vertical gradient measurements with and without the cover and a radial measurements only at Z=50 cm with cover in place. Each measurement lasted 1200 s, divided into 240 acquisitions of 5 s length each. Figure 10 shows a 3D counts map recorded at Z=50 cm, elaborated by means of a MATLAB routine.

The GEM detector allows at obtaining the neutron flux map inside the HOTNES cavity. In figure 11, are shown the four colour maps at different planes (Z=30, 40, 50 and 60 cm), which highlight the good homogeneity level of thermal neutron flux for Z=40, 50 and 60 cm planes, as shown from TNPD measurements.

The quantity q(Z,Σ), i.e. the vertical gradient of the GEM count rate but measured over the whole pads area (Σ), shows a trend that well resembles those shown before for the other detectors (see figure 12).

An experimental value of the efficiency, ε, can be obtained by averaging over the determination of ε obtained at the different planes investigated, then providing a value of ε≅0.011. This value is in satisfactory agreement with the prediction of Monte Carlo simulations and on analytical calculations based on the effective $^{10}$B thickness that the GEM cathode features [22].

## 2.5. Pure Caesium Iodide crystal calorimeter

The Mu2e experiment at Fermilab aims to measure the neutrinoless conversion of a negative muon into an electron, reaching a single event sensitivity of $2.5 \times 10^{-17}$ after three years of data taking [23], thus improving of four orders of magnitude the previous measurement. The mono-energetic electron produced in the final state is detected by a high precision tracker with 120 keV momentum resolution and a crystal calorimeter, all embedded inside a large detector solenoid surrounded by a cosmic ray veto system. The calorimeter is complementary to the tracker, allowing an independent trigger, powerful particle identification and seeding for track reconstruction. In order to match these requirements, the calorimeter should have an energy resolution of ~ 5% and a time resolution better than 500 ps at 100 MeV [24]. The choice of the calorimeter components went through an intense R&D phase. The selected option is a calorimeter composed of two disks of pure Caesium Iodide (CsI) crystals [25], 34×34×200 cm$^3$, read by UV-extended Silicon PhotoMultipliers (SiPM) [26].

An important issue for the Mu2e calorimeter, related to the high radiation environment, is that of controlling the phosphorescence light induced by radiation and evaluate the corresponding equivalent energy noise when the crystals are irradiated with a neutron fluence or an ionization dose similar to the one expected while running in Mu2e. The Radiation Induced Noise (RIN) for CsI crystals has been evaluated at the HOTNES facility on six different samples, produced by ISMA/AMCRYS (Ukraine), Opto Materials (Italy) and SICCAS (China) companies. A single crystal, coupled to a photomultiplier with a gain of $2.1 \times 10^6$ at 1400 V, has been inserted inside the HOTNES cavity and thus irradiated with a uniform thermal neutron flux, $\phi_n \sim 700$ cm$^{-2}$ s$^{-1}$. The dark current values, $I_{dark}$, have been recorded by means of a Keythley and an automatic acquisition program. To evaluate the radiation induced noise we are first interested in determining F, which is the induced number of photoelectrons ($N_{pe}$) per second for the given neutron flux, defined as: $F = I_{dark}/(\varepsilon \times G_{PMT}) \times \phi_n$. To evaluate the noise expected for the Mu2e experiment we considered a readout gate of 200 ns and an estimated flux of $10^4$ cm$^{-2}$ s$^{-1}$.

In these conditions, the number of radiation induced photoelectrons is derived as: $N_{pe} = F \times \phi_{Mu2e} \times 200$ ns. Finally, from this number, the RIN can be obtained using the following relation: $RIN = N_{pe}^{0.5}/LY$, where LY is the light yield of the crystal. Results are reported in table 2, while in figure 13 examples of signals obtained for crystals of the three different vendors are shown.

The value of the current increase from $I_{dark} \sim 10$ nA while the crystal is not irradiated to $I_{neutrons} \sim O(10)$ µA under irradiation. All the crystals tested showed a similar behavior and the RIN values obtained, that is between 60-85 keV for a flux of $10^4$ cm$^{-2}$ s$^{-1}$, meet the Mu2e requirement of a RIN < 0.6 MeV.

### 2.6. Single Crystal Diamond detector

Single crystal diamond detector have shown their effectiveness in detecting fast neutrons in different neutron environments, such as fusion experimental machines [27], spallation neutron sources for time of flight fast neutron counting [28, 29]. These detectors are well suited for high temperature neutron measurements [30, 31].

Single crystal diamond detectors featuring a Lithium Fluoride ($^6$LiF) coating are under test for future application to the ITER experiment [32].

The prototype detector was designed by ENEA and is schematically sketched in figure 14. The detector layout is based on mechanical connections to ground and HV electrodes. The use of a mineral cable (MI) (1m long in the present prototype) and a ceramic connector (MOC) also allow high-temperature operation. These two elements can withstand temperatures of up to 800 °C and 400 °C, respectively, as well as intense radiation fluence [33]. The diamond film ("electronic grade" type) had a 4.3 × 4.3 mm$^2$ area and was 500 µm thick, and was produced by e-6 Ltd. [34]. It was first annealed at 500 °C for 1 h in Ar and then one Ag metal electrode, 100 nm thick, was deposited by thermal evaporation on each surfaces. A 2 µm thick $^6$LiF layer was then deposited by thermal evaporation on top of one of the two Ag contacts, thus making the detector sensitive to thermal neutrons through the nuclear reaction n+$^6$Li → $^4$He + $^3$H . Both $^4$He and $^3$H particles can be detected by the diamond detector.

Before using this device on beam at the ISIS spallation neutron source, it was tested on the HOTNES facility to investigate its response to thermal neutrons, both in terms of amplitude as well as in terms of time characteristics. The experimental setup was made of a CAEN digitizer (Model 5730) able to register waveforms, obtained by passing the SCD signal to a fast preamplifier. A trapezoidal filter implemented was used to integrate the charge and thus enabling to construct pulse height spectra.

Figure 15 shows a biparametric (ADC-channel vs time) plot that is used to assess the stability of the device over a given recording time.

After 5000 s of acquisition some noise was registered at low ADC channel. At higher channel the spectrum is clean as it is shown in figure 15 where the signal from the n-$^6$Li reaction is visible. Figure 16 represents the projection of the biparametric plot (integrated over the whole acquisition time) shown in figure 15 onto the ADC-channel axis and then represents the pulse height spectrum.

Figure 17 shows a typical pulse registered by the digitizer when operated in the oscilloscope mode.

These measurements were done in preparation of the experimental campaign performed at the ISIS spallation source to test the same acquisition electronics for high-resolution neutron measurements that will be discussed in a forthcoming experimental paper [35].

### 3. Conclusion and perspectives

The results of the first HOTNES user programme were reported.

Different detectors were tested, namely Li-glass scintillation detector, Cadmium-Zinc-Telluride semiconductor detector, Thermal Neutron Rate Detectors (TNRD), Single Crystal Diamond detector featuring a Lithium Fluoride coating and a neutron-sensitive Gas Electron Multiplier.

The following characteristics made the campaign successful:
- The reduced gamma background (kerma rate 4-8 µGy/h, according to the irradiation plane) allowed testing those sensor that are also sensitive to gammas;
- The large irradiation volume allowed introducing in HOTNES devices with bulky equipment such as PM tubes or electronics modules;

- The very high homogeneity in the thermal field across a given irradiation plane allowed exposing multiple detectors at the same time or devices with large sensitive area (like the GEM);
- The thermal flux, in the order of 700-1000 $cm^{-2} s^{-1}$ (according to the irradiation plane) was large enough to successfully test detectors that are normally used in fields with high-fluence rate (like the TNRD).

Although HOTNES can be considered already a complete tool for testing neutron sensitive devices, further improvements are planned: determining the photon spectrum, and equipping the facility to irradiate personal neutron dosemeters on phantom according to ISO Standards [36].

### Acknowledgments
The HOTNES facility was set up in the framework of ENEA / INFN-LNF collaboration.

**Figure Captions**

**Figure 1**. Frontal and top views of HOTNES.

**Figure 2**: CAD model of HOTNES. Dimensions are expressed in centimeters.

**Figure 3**: Pulse height spectrum recorded by the Li-glass scintillation detector placed at the Z=50 and R=0. The arrow indicated the signal from the n-$^6$Li reaction (Q=4.78 MeV). The photon signal is below channel 50.

**Figure 4**: Trend of q(Z,R) measured by the Li-glass detector without HOTNES polyethylene cover in place.

**Figure 5**: Trend of q(Z,0) measured by means of the Li-glass compared with the one measured by means of TNPD.

**Figure 6**: Pulse height spectrum recorded by the CZT detector placed at Z=50 and R=0. The peak around ADC-Channel 705 is the 558.46 keV thermal neutron radiative capture gamma-ray on $^{113}$Cd.

**Figure 7**: Trend of q(Z,R) for the CZT detector with the polyethylene cover in place.

**Figure 8**: Trend of q(Z,0) measured by means of the CZT and compared with the one measured by means of TNPD with the cover in place.

**Figure 9**: Typical TNRD output voltage during the three stages of the calibration in HOTNES.

**Figure 10**: 3D representation of GEM counts distribution over the whole set of pads (labeled by the integers in the axes) relative to plane Z=50 cm.

**Figure 11**: 2D maps showing the neutron field intensity distribution as number of events above threshold recorded at each GEM readout pad for Z=30, 40, 50 and 60 cm.

**Figure 12**: Trend of q(Z,Σ) measured by means of the GEM and compared with the one measured by means of TNPD without the cover. See text for details.

**Figure 13**: (From left to right) Example of signal for Opto Materials, SICCAS and ISMA crystals under test to evaluate the radiation induced noise.

**Figure 14**: Schematic of the SCD detector tested at the HOTNES facility.

**Figure 15**: Scatter plot ADC-channel vs recording time. The signal at channel 1500 are events from n-$^6$Li reactions. After 1500 s, a noise pickup is present at low ADC channels (below channel 700).

**Figure 16**: Pulse height spectrum obtained by projecting the scatter plot in figure 15 on the ADC channel axis and integrating over the whole time window.

**Figure 17**: Typical waveform from a n-$^6$Li event, registered by the CAEN digitizer when operated in the oscilloscope mode.

**Tables captions**

**Table 1**. Calibration factors of six TNRDs.

**Table 2**: Results from RIN test at the HOTNES Facility.

**Figures**

Figure 1

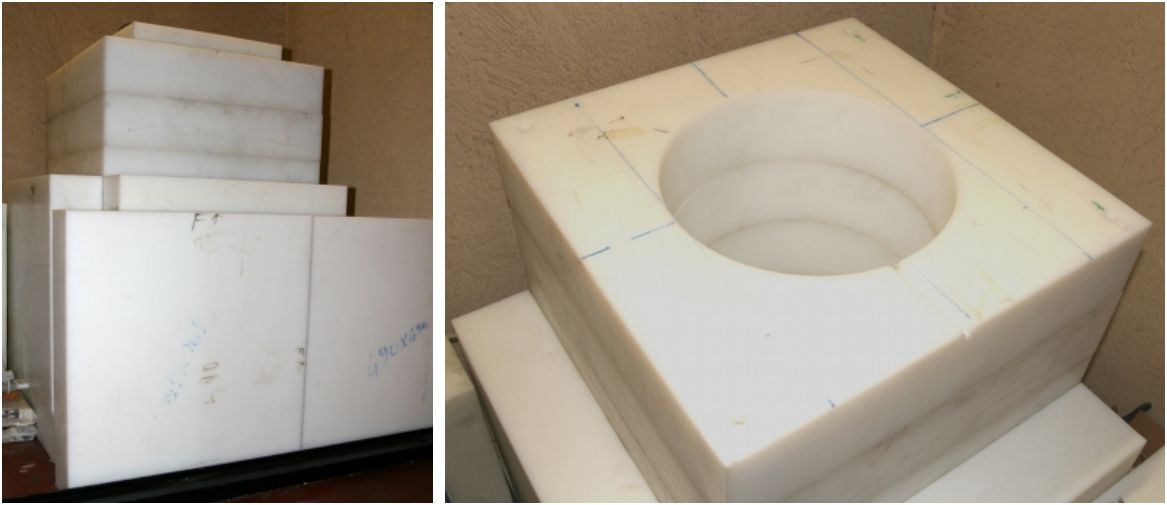

Figure 2

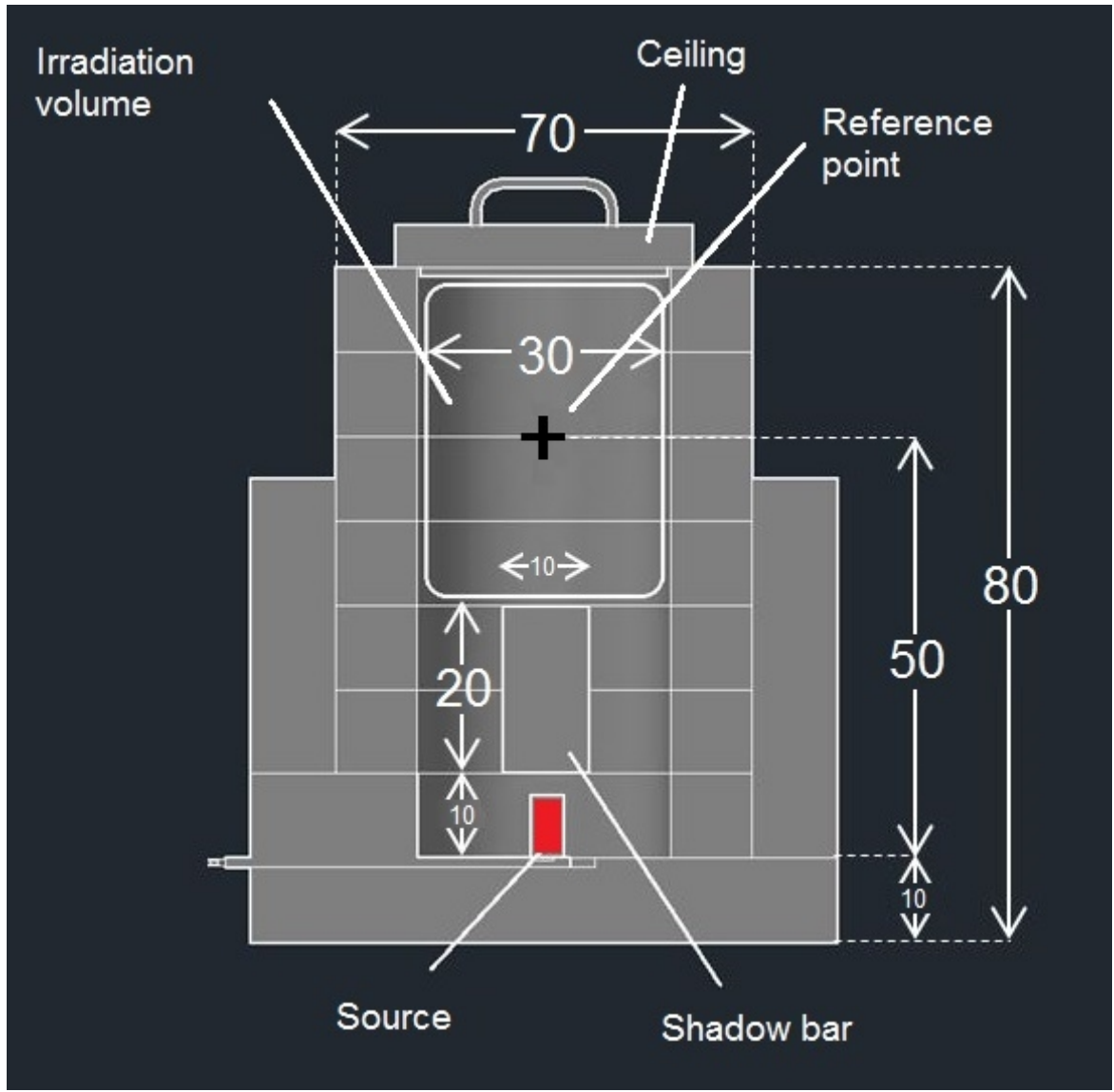

Figure 3

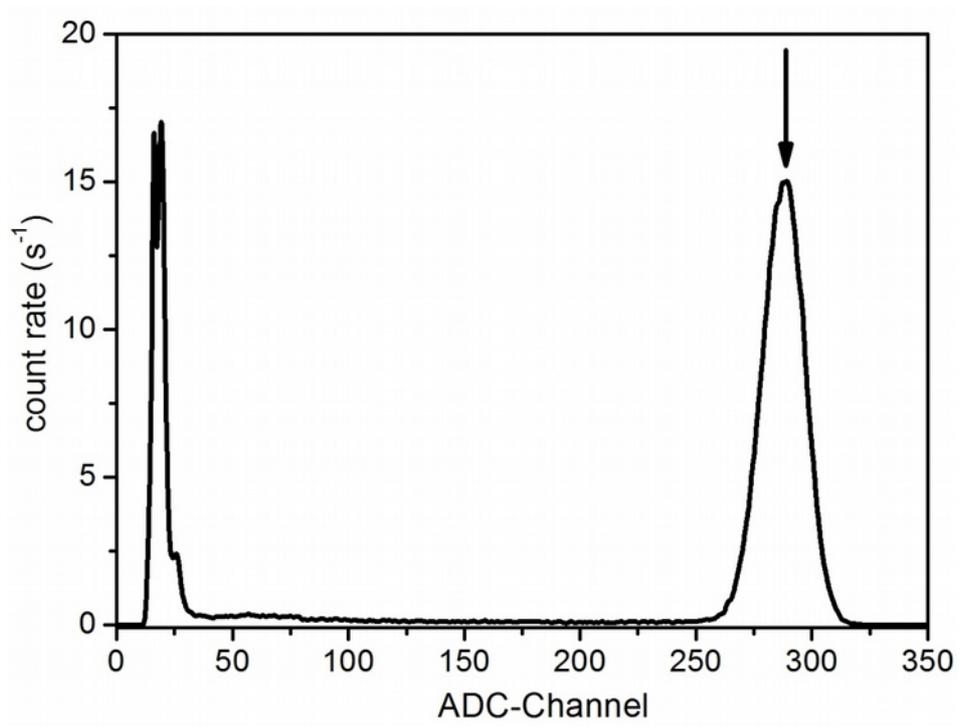

Figure 4

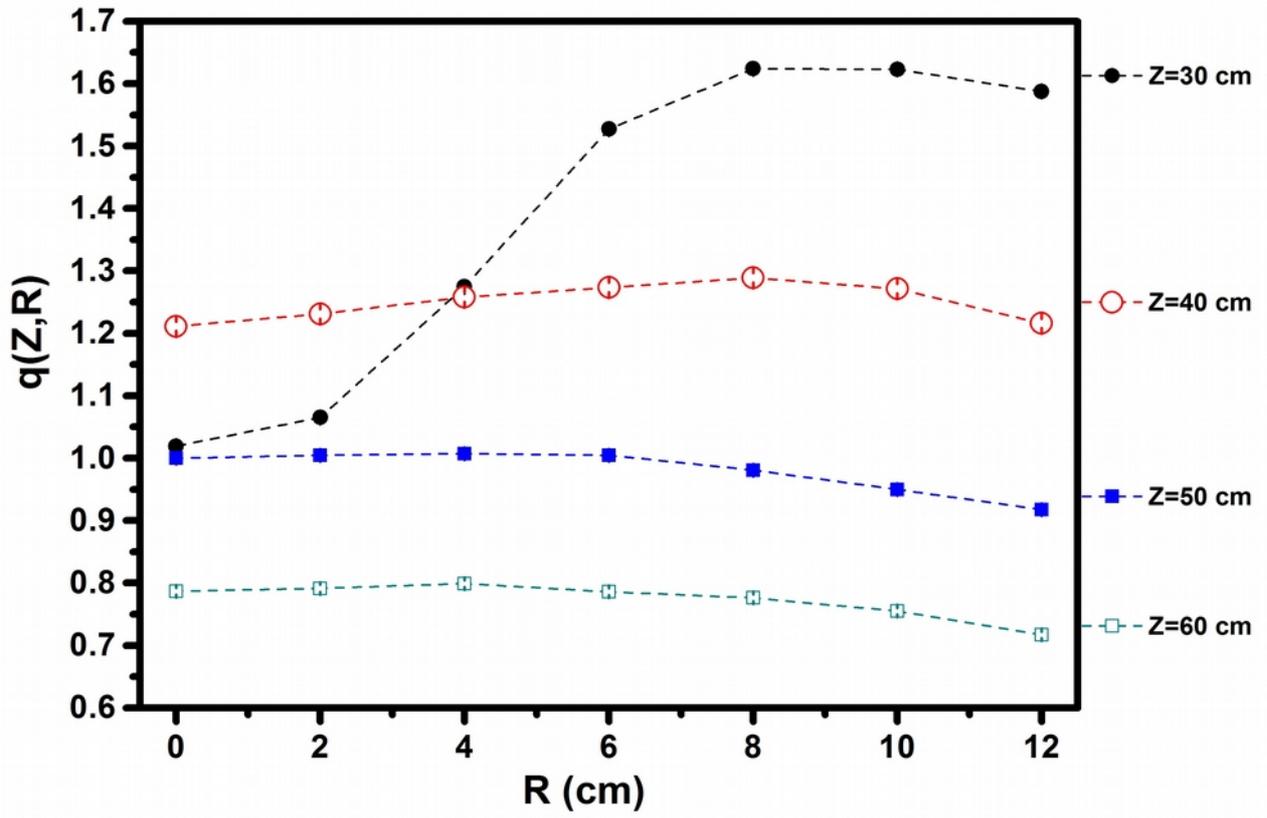

Figure 5

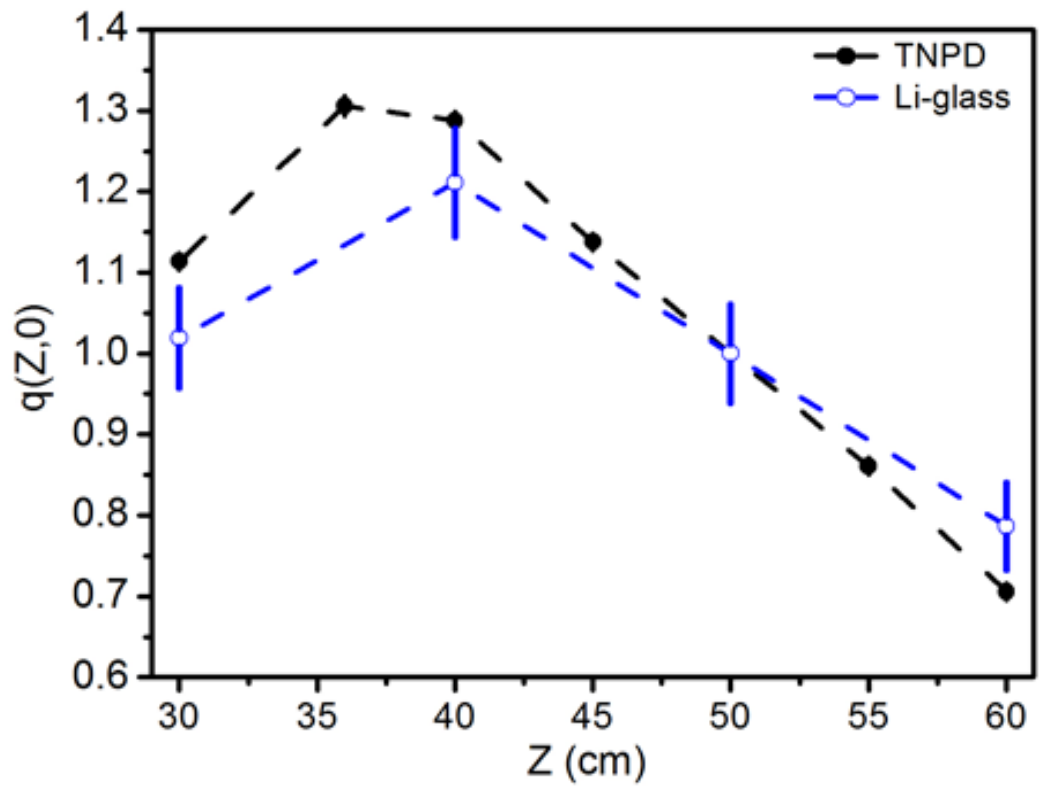

Figure 6

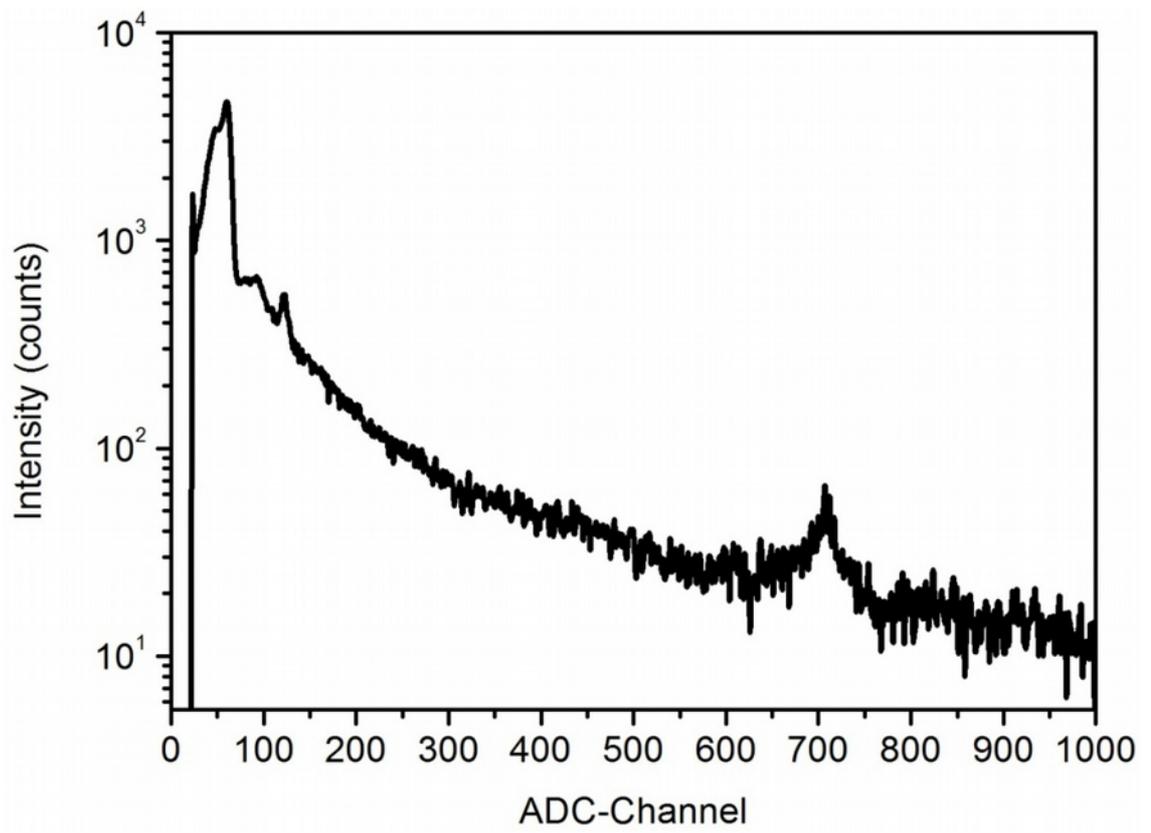

Figure 7

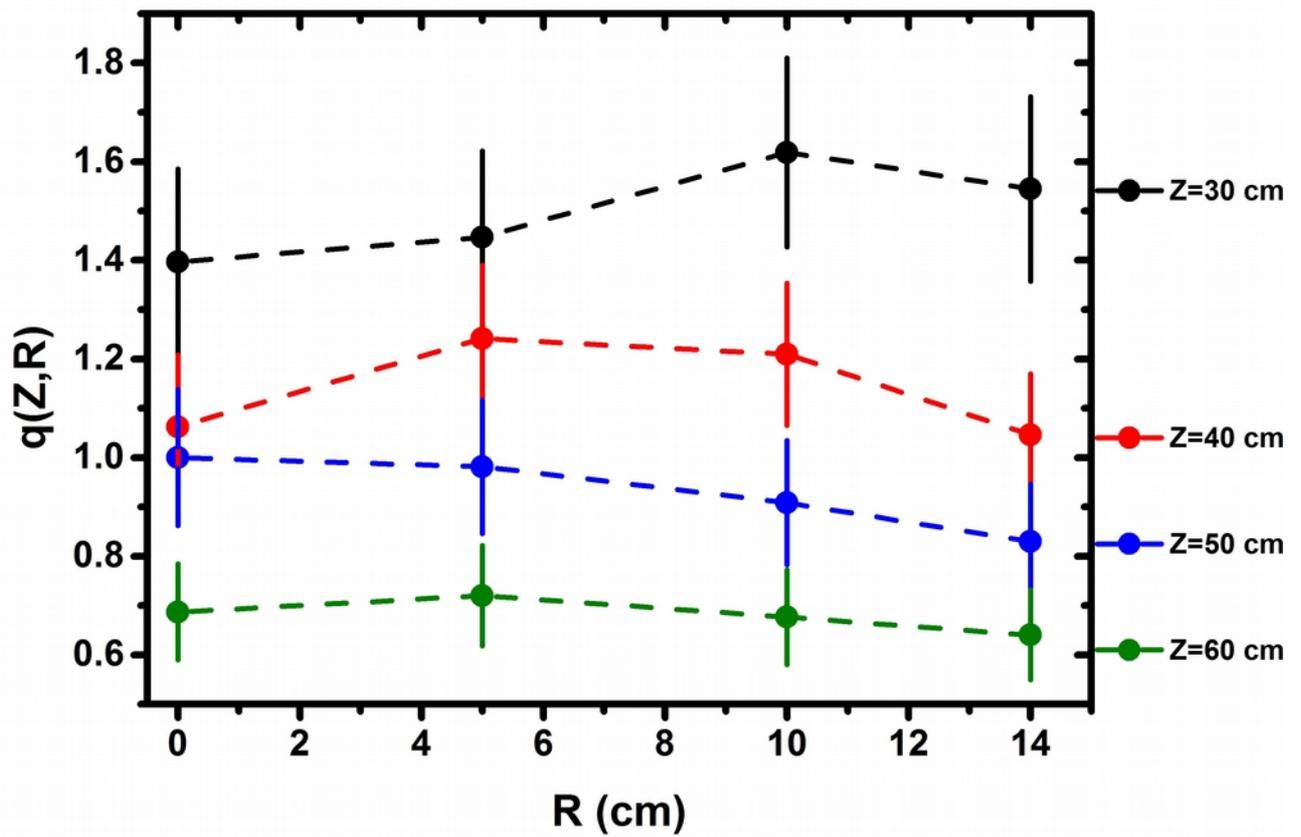

Figure 8

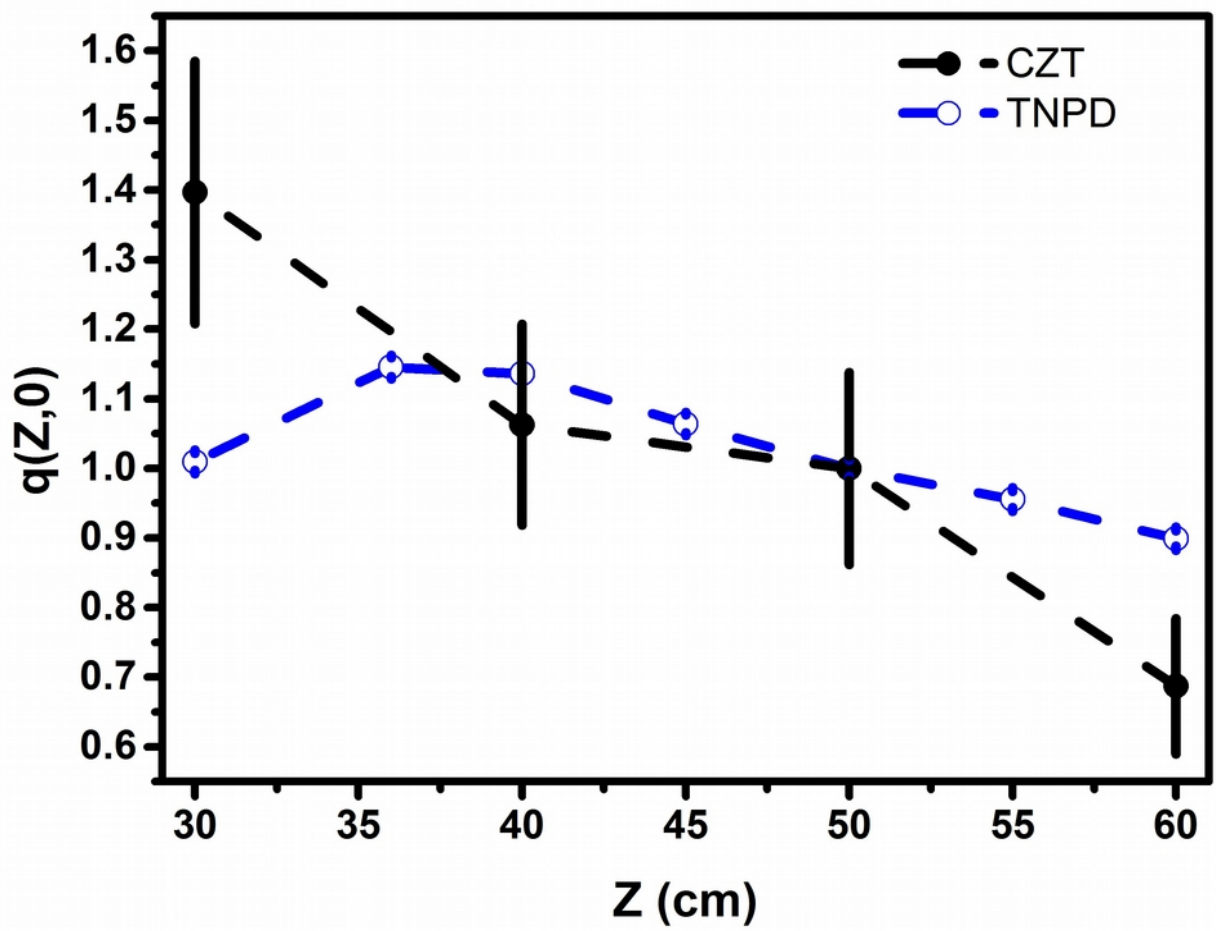

Figure 9

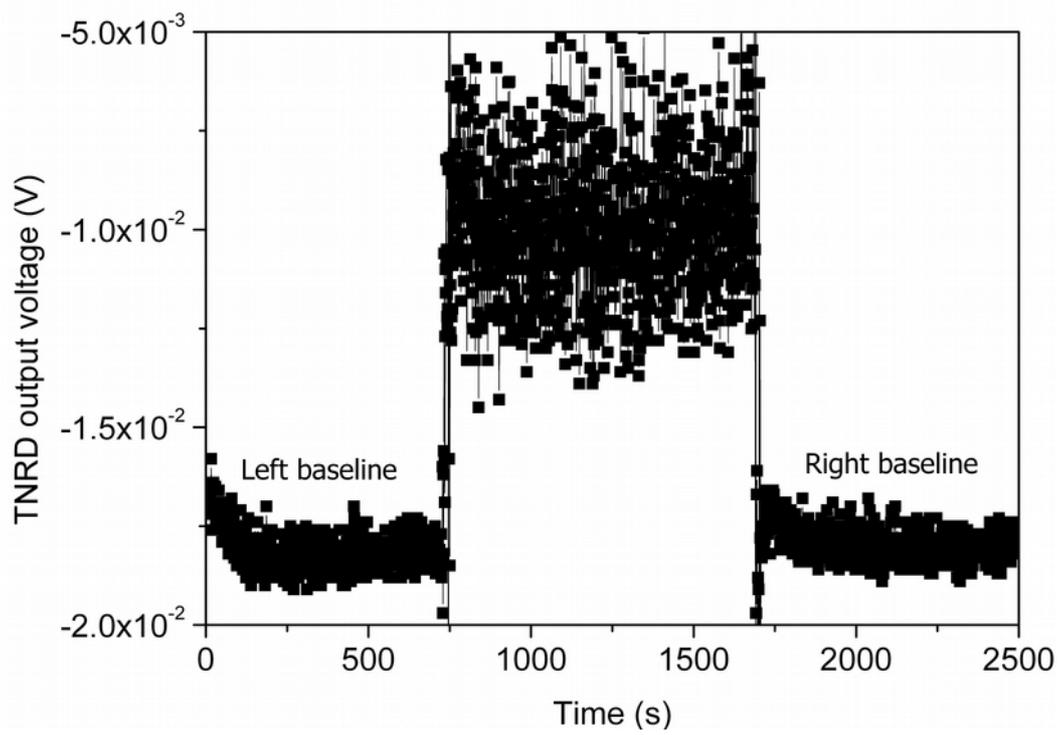

Figure 10

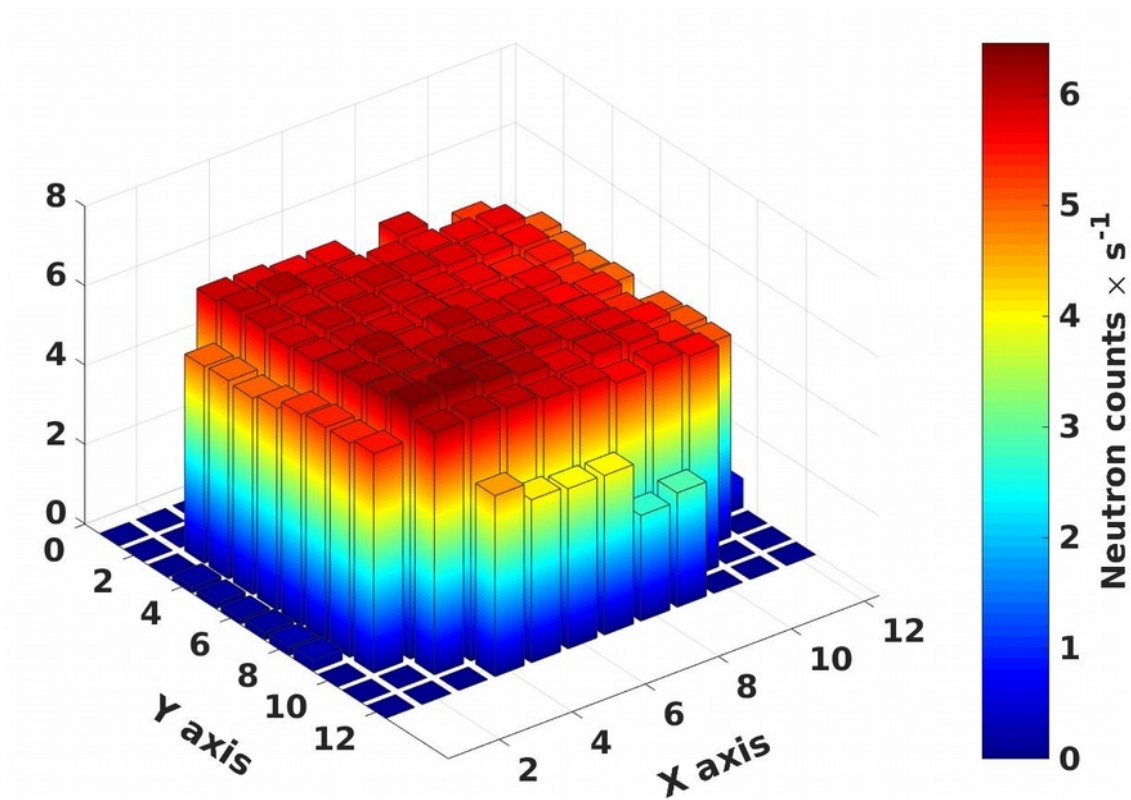

Figure 11

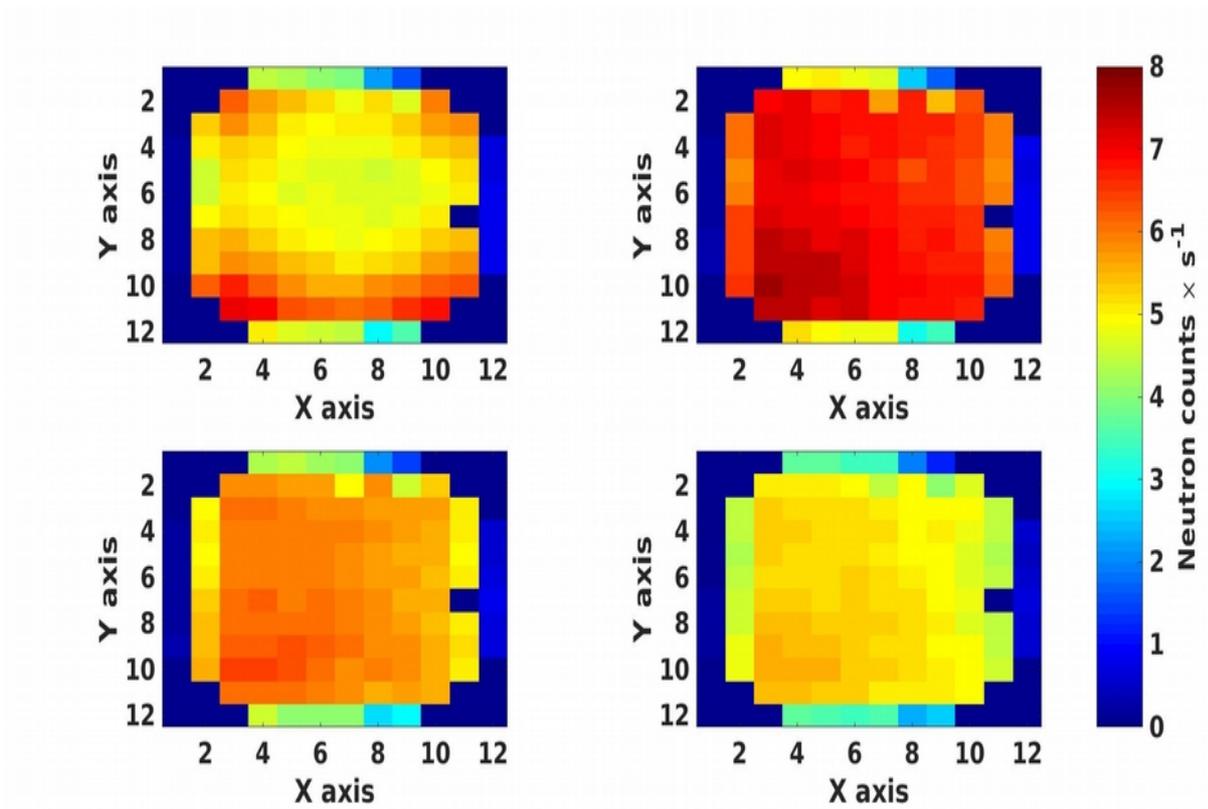

Figure 12

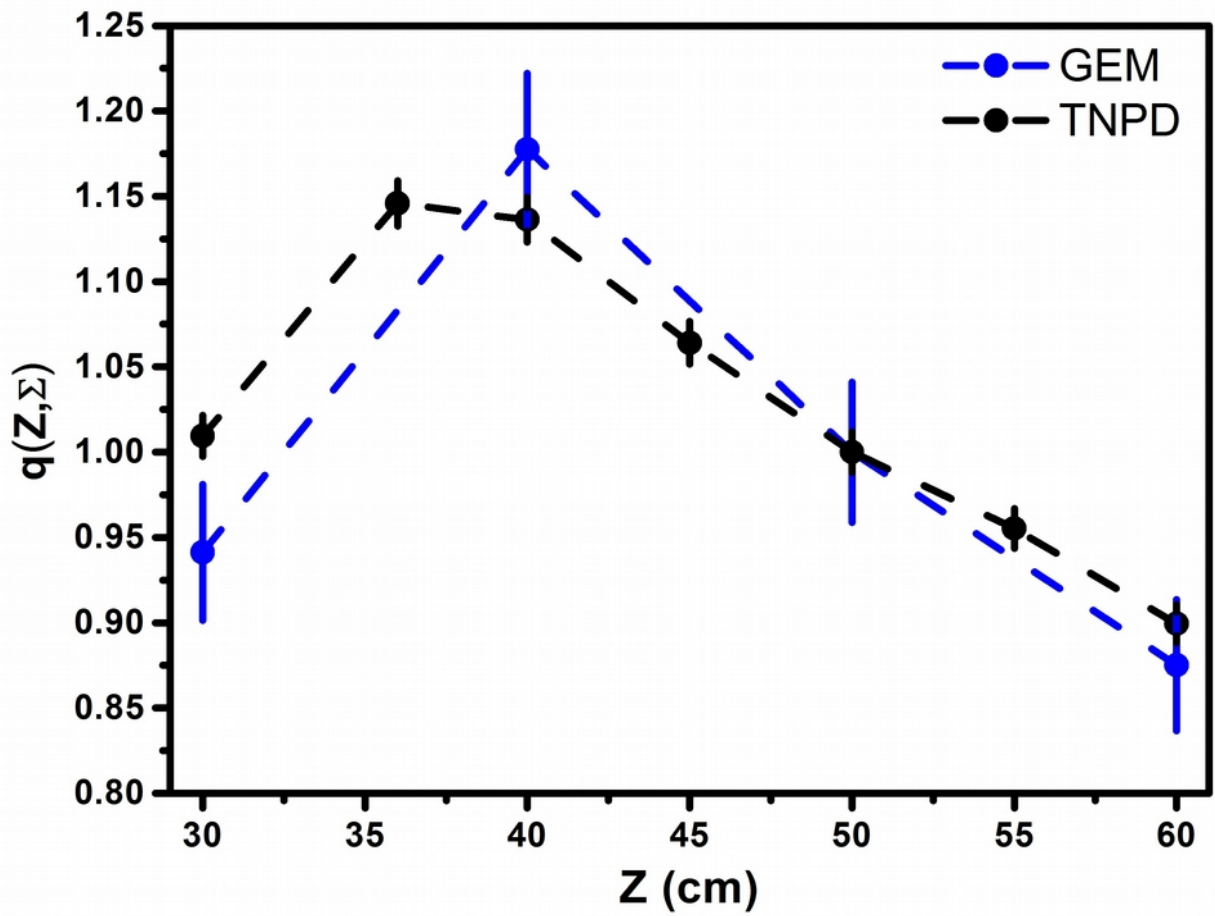

Figure 13

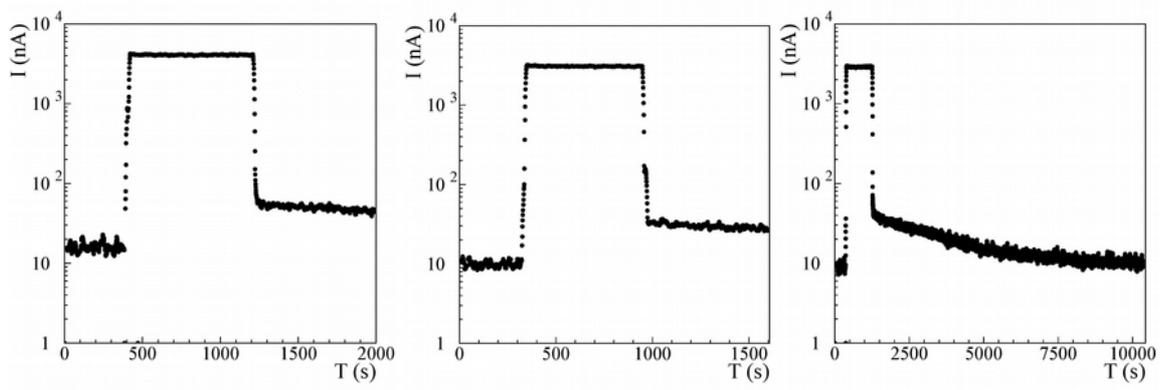

Figure 14

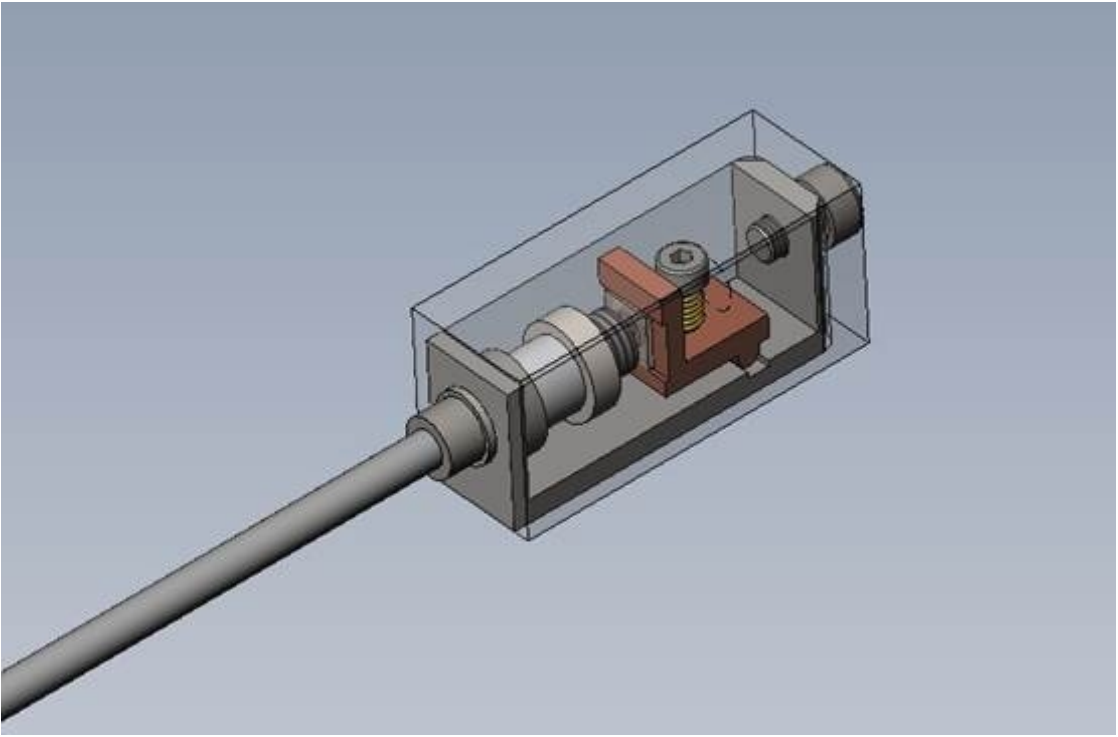

Figure 15

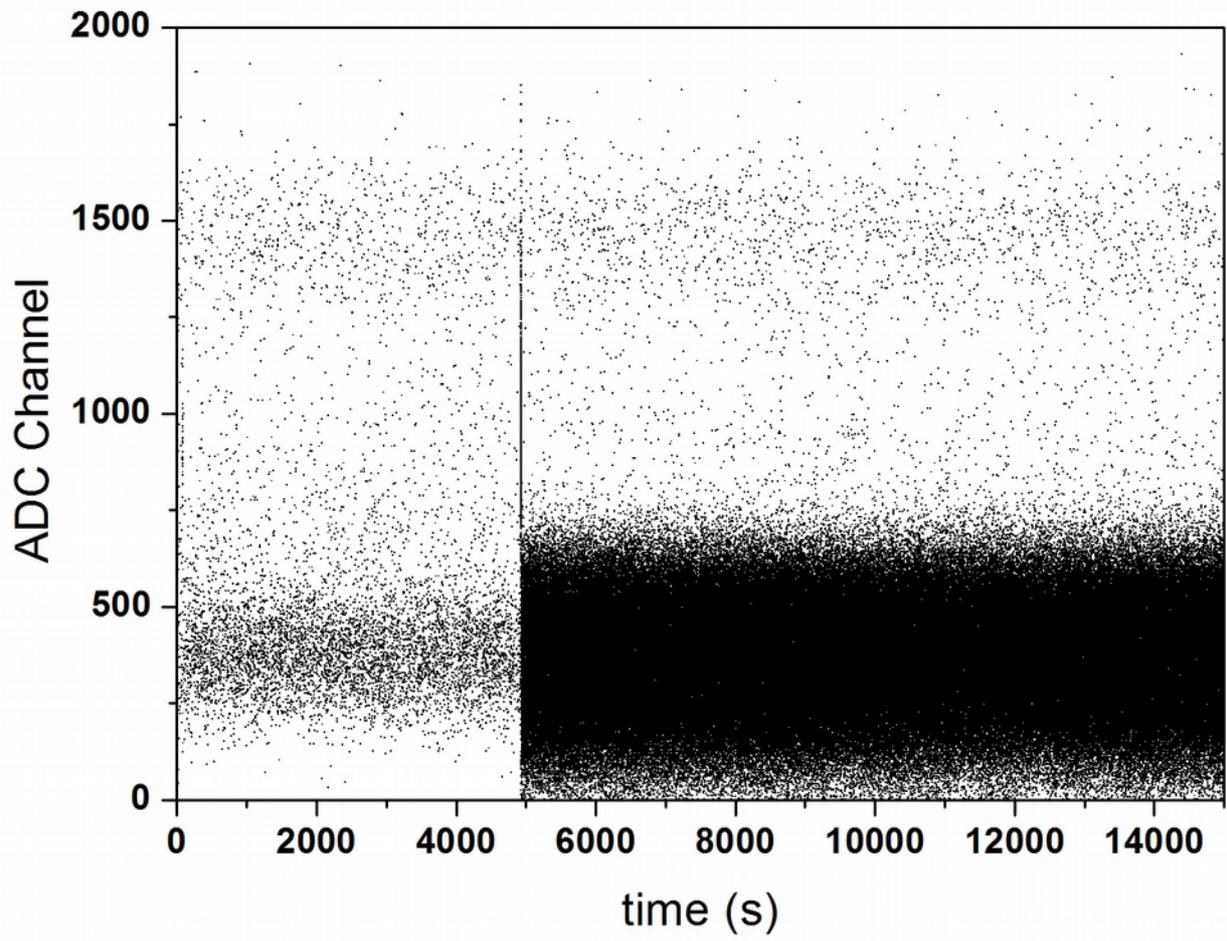

Figure 16

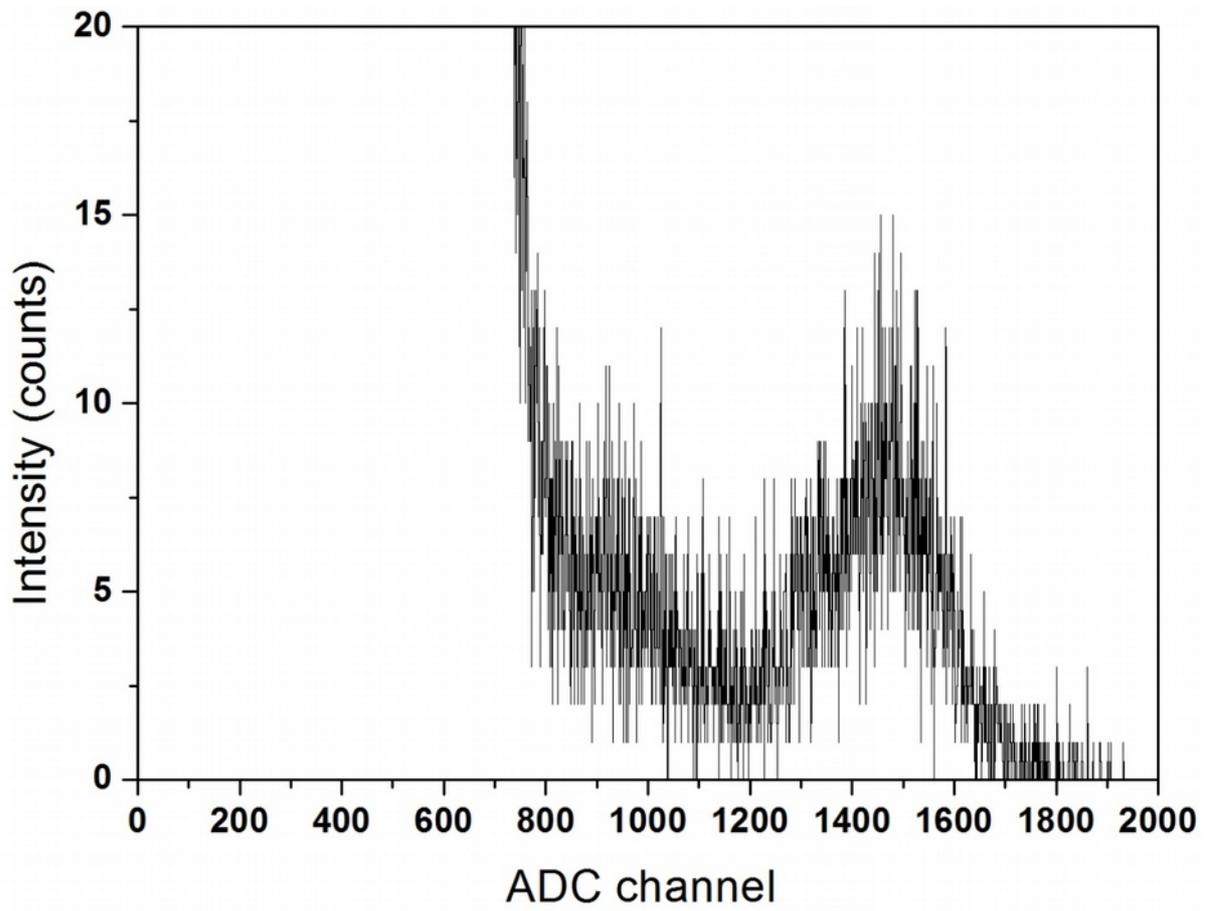

Figure 17

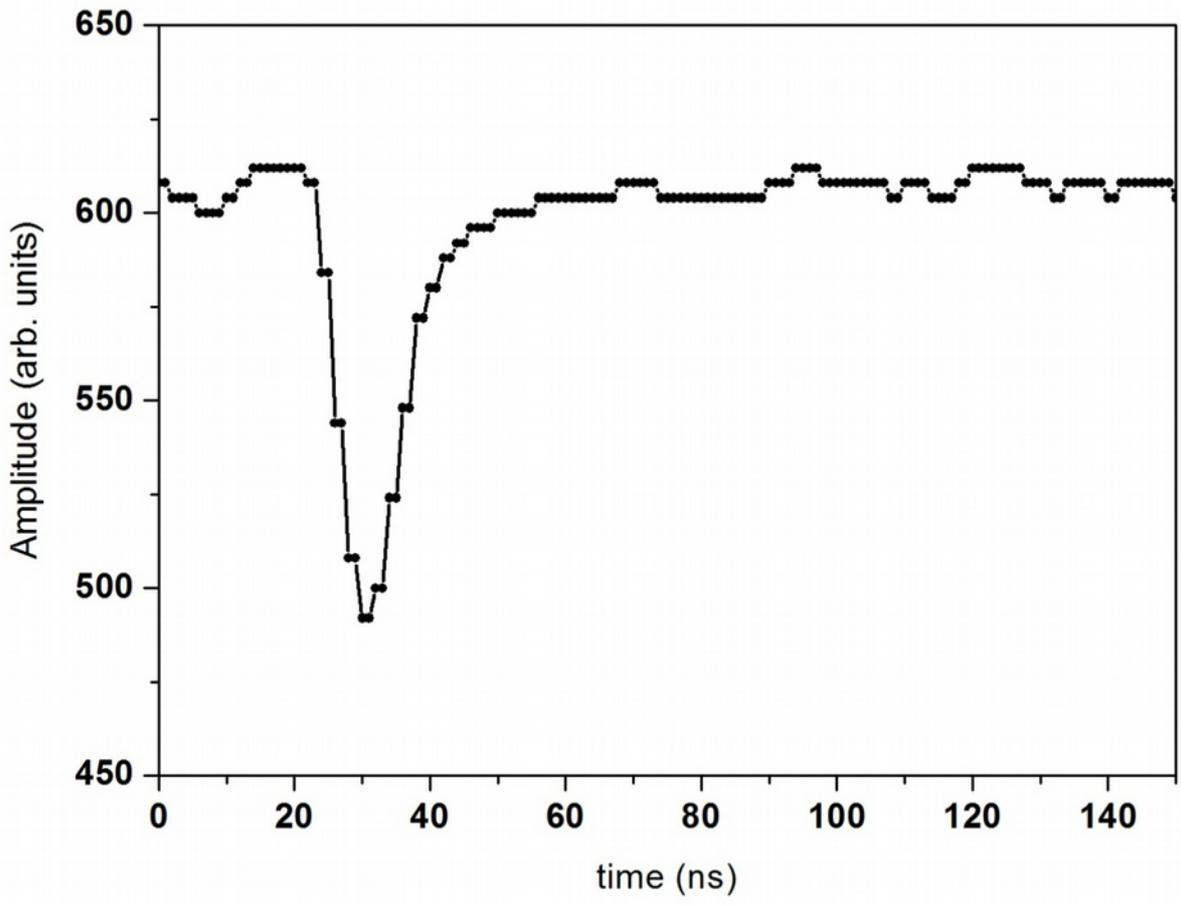

**Tables**

**Table 1**

| TNRD code | Calibration factor ($\mu$V cm$^2$ s) |
|---|---|
| 1 | 13.1±0.5 |
| 2 | 11.8±0.5 |
| 3 | 10.9±0.4 |
| 4 | 10.3±0.4 |
| 5 | 10.8±0.4 |
| 6 | 13.4±0.5 |

**Table 2**

| Crystal | LY [$N_{p.e.}MeV^{-1}$] | $I_{dark}$ [μA] | F [$N_{p.e.}s^{-1}\phi_n^{-1}$] | $N_{p.e.}$ | $RIN_{Mu2e}$ [keV] |
|---------|---------|---------|---------|---------|---------|
| ISMA 02 | 103 | 7.16 | $3.02 \times 10^4$ | 60.3 | 75.4 |
| ISMA 12 | 103 | 4.61 | $1.94 \times 10^4$ | 38.9 | 60.5 |
| ISMA 20 | 103 | 5.35 | $2.25 \times 10^4$ | 45.1 | 65.2 |
| ISMA 21 | 103 | 7.28 | $3.07 \times 10^4$ | 61.4 | 76.0 |
| SICCAS 1 | 129 | 6.83 | $2.88 \times 10^4$ | 57.5 | 58.6 |
| SICCAS 2 | 126 | 7.58 | $3.19 \times 10^4$ | 63.8 | 63.4 |
| SICCAS 4 | 136 | 10.1 | $4.27 \times 10^4$ | 85.5 | 67.8 |
| OPTOM 2 | 93 | 7.65 | $3.22 \times 10^4$ | 64.4 | 86.3 |